\documentclass[twocolumn,showpacs,preprintnumbers,amsmath,amssymb,prl]{revtex4}
\usepackage{graphicx}
\usepackage{dcolumn}
\usepackage{bm}
\usepackage[FIGTOPCAP, hang, nooneline]{subfigure}
\usepackage[usenames,dvipsnames]{color} 
\usepackage{mathrsfs}
\usepackage[english]{babel}
\usepackage[latin1]{inputenc}
\usepackage{blindtext}

\newcommand{\uto}[1]{{\ }^{\underrightarrow{\ #1 \ }} \, }

\newcommand{\oo}{\varnothing} 
\newcommand{\dm}[1]{begin{displaymath} #1 \end{displaymath}}
\newcommand{\eq}[1]{begin{equation} #1 \end{equation}}

\newcommand{\bdm}{\begin{displaymath}}
\newcommand{\edm}{\end{displaymath}}
\newcommand{\be}{\begin{equation}}
\newcommand{\ee}{\end{equation}}
\newcommand{\ba}{\left( \begin{array}}
\newcommand{\ea}{\end{array} \right)}

\newcommand{\tn}{\widetilde{n}}

\begin{document}

\title{Coexistence in  a One-Dimensional Cyclic Dominance Process}

\author{Anton A. Winkler$^1$} \author{Tobias Reichenbach$^2$} \author{Erwin Frey$^1$}
\affiliation{$^1$Department~of~Physics, Ludwig-Maximilians-Universit\"at M\"unchen, Theresienstra\ss e
 37, 80333 M\"unchen, Germany \\
$^2$Howard Hughes Medical Institute and Laboratory of Sensory Neuroscience, The Rockefeller University,
1230 York Avenue, New York, NY 10065, U.S.A.
 }

\date{\today}

\begin{abstract}
Cyclic (rock-paper-scissors-type) population models serve to mimic complex species interactions. Focusing on a paradigmatic three-species model with mutations in one dimension, we observe an interplay between equilibrium and non-equilibrium processes in the stationary state. We exploit these insights to obtain asymptotically exact descriptions of the emerging reactive steady state in the regimes of high and low mutation rates. The results are compared to stochastic lattice simulations. Our methods and findings are potentially relevant for the spatio-temporal evolution of other non-equilibrium stochastic processes.
\end{abstract}


\maketitle

Stochastic interacting particle systems are a fruitful testing ground for understanding generic principles in non-equilibrium dynamics.
Unfortunately, the treatment of such processes is marred by the absence of detailed balance so that the insight one has gained by analytical means is not yet satisfactory and only few systems have been solved exactly \cite{privman, liggett2}.
Some of them serve as a paradigm for very complex biological and sociological systems. An example is the contact process, which describes the outbreak of an epidemic, displaying a phase transition from an 
absorbing to an active state as the rate of infection is increased \cite{hinrichsen}. Another famous process is the voter model, caricaturing opinion making. 
It is proven rigorously that on a regular lattice there is a stationary state where the two ``opinions" coexist, so long as the dimension is larger than 2, such that the random walk is not recurrent \cite{Castellano:2009p8700,liggett2}. Extensive studies have also been conducted on the coarsening dynamics of coalescing or annihilating particles, both for diffusional motion and ballistic motion of the particles \cite{bray,frachebourg-1996-77,frachebourg,kafri}. In this context, much work was devoted to the long time behavior of the average domain size, which as a function of time typically displays scaling. 

Frachebourg et. al. \cite{frachebourg-1996-77,frachebourg} have studied the coarsening dynamics of a model known as the Rock-Paper-Scissors game (RPS),
one of the most widely studied prototype models for biodiversity \cite{kerr, reichenbach-2006-74, reichenbach-2007-448}, displaying cyclic dominance between its three agents.
In this Letter we study the influence of mutations on this model.
An integral part of evolution, mutations have been posited to promote biodiversity in microbial communities \cite{Czaran:2002p3816}. We will argue that the RPS is a natural framework for a nonequilibrium version of the Ising-Glauber model, which at zero temperature amounts to an annihilating random walk.
While previous studies have addressed coarsening and the transition to an absorbing state, we focus on the description of the stationary reactive state at finite ``temperature", i.e. interfaces between domains are created at finite mutation rate. In the Ising-Glauber model the interfaces perform a random walk, whereas for the RPS they drift left or right and even move ballistically in a certain regime. Since the coarsening dynamics is counteracted by the creation of interfaces, the system evolves into a non-trivial stationary state. For very large and very low mutation rates, equilibrium turns out to be only slightly broken. Discriminating between two types of mutations, we can thus obtain asymptotically exact descriptions for the average size of the domains in the stationary state. As the final arbiter of the validity of our arguments we employ stochastic lattice simulations.

\begin{figure} 
\centering
 \includegraphics[width=0.47\textwidth] {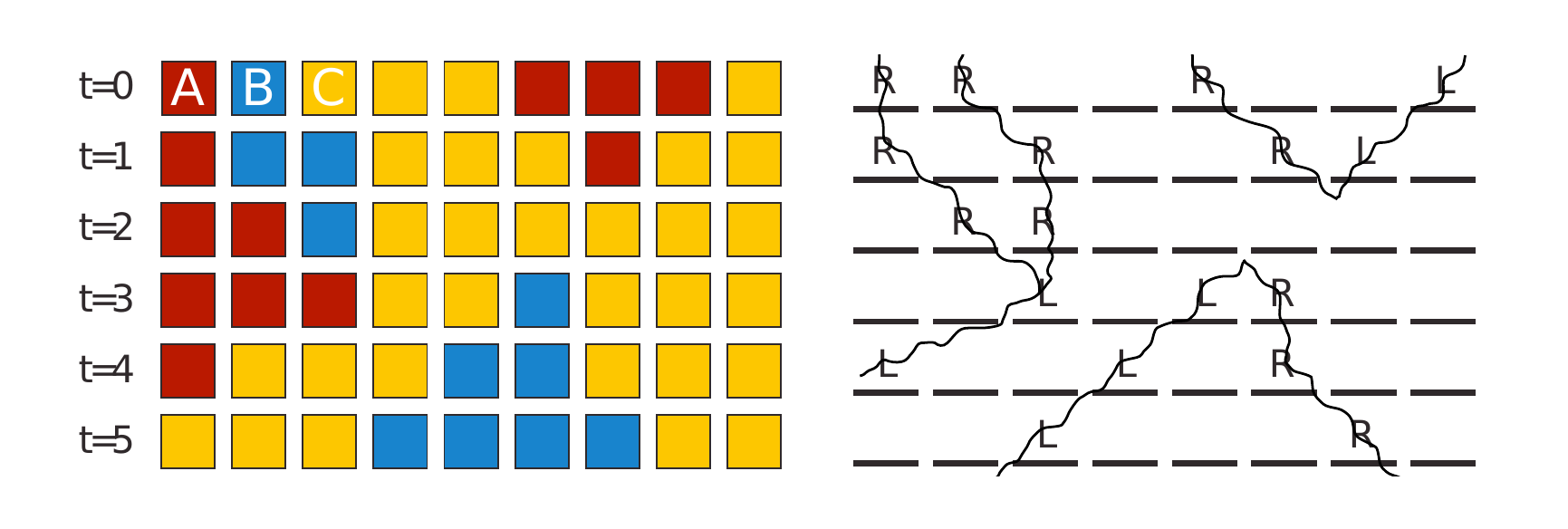}
 \caption{Illustration of the one-dimensional Rock-Paper-Scissors game with mutations and the passing to its dual description. The configuration of the lattice is given at subsequent points in time, resulting in a two-dimensional space-time plot. A mutation $C \to B$ occurs somewhere between $t=2$ and $t=3$. The dual picture is characterized by interfaces moving left ($L$) or right ($R$). 
}
 \label{fig:pictogram}
\end{figure}

On a one-dimensional integer lattice $\{1,\ldots,S\}$ of size $S$, the RPS can be defined by the following cyclic-dominance reaction equations for nearest neighbors.
\begin{equation}
 AB  \uto{r_A} AA \,, \quad  BC \uto{r_B} BB \,, \quad CA \uto{r_C} CC \,, 
\label{eq:LotkaVolterra}
\end{equation}
i.e.~paper ($A$) covers rock ($B$), rock crushes scissors ($C$) and scissors cuts paper.
Here we presuppose left-right symmetry such that, for instance, $A$ can invade a neighboring $B$ to its left or right and we consider a Markov process in continuous time with sequential updating.
Unless otherwise stated, we look at the symmetric case $r_A = r_B = r_C$ and set $r_A = 1$ to define the timescale. These equations have been studied in detail in~\cite{frachebourg-1996-77,frachebourg}. In particular it was shown that, starting from some random distribution, the species organize in domains that 
undergo coarsening until finally---providing the system size is finite---one species takes over the whole lattice.

In addition to the above reaction scheme for cyclic dominance we allow for mutations, 
\begin{equation}
 A \uto{\mu_r} B \uto{\mu_r} C \uto{\mu_r} A \,,\quad   A \uto{\mu_l} C \uto{\mu_l} B \uto{\mu_l} A \,, 
\end{equation}
where we discriminate mutation cycles to the respective ``prey'' with rate $\mu_r$ and to the respective ``predator'' with rate $\mu_l$, both of which evidently conserve the cyclic symmetry.
The mutations counteract the coarsening of the system and ensure a reactive steady state. However, for low mutation rates, which we shall focus upon, one still expects the system to organize in large clusters separated by interfaces. Thus it is adequate to utilize the so-called dual description, obtained by representing the interfaces, i.e.~the walls between the domains of one species, by particles (denoted $L$ or $R$) and two consecutive spots occupied by the same species by empty sites $\oo$. This mapping is illustrated in Figure~\ref{fig:pictogram}. 
There are left- and right-moving interfaces, $R$ and $L$ respectively. Without mutations their number 
is bound to decrease when they interact: 
in $LL \uto{1} R \oo$ (and analogous for left and right interchanged, $RR \uto{1} \oo L$) two interfaces of the same kind turn into one that moves in the converse direction and in $RL \uto{2} \oo \oo$ one has pair annihilation. 

As a starting point we derive the mean-field rate equations. Let $P(R), P(L), P(\oo)$ be the probabilities of finding an $R$, $L$ or $\oo$, respectively, at some site $i$. One then assumes the system to be well mixed, i.e. the occupancy of sites to be uncorrelated. Defining $\mu := \mu_l + \mu_r$, one obtains, 
	$\dot P(R) = - 2 P(R)^2 - 2 P(R) P(L) + P(L)^2 + \mu \left(P(\oo) + P(L) - 2 P(R)\right) $.  Solving for the stationary state, where $P(R) = P(L)$, the interface density becomes
\begin{equation}
	n := P(R) + P(L) = \sqrt{(4/3) \mu + \mu^2} - \mu\, \,,
	\label{eq:mean-field}
\end{equation} 
increasing sharply $\propto  \sqrt{\mu}$ for small rates and saturating to $2/3$ for large rates.
Notice that $\mu_r$ and $\mu_l$ are treated on equal footing in contrast to our results below.

Before we proceed with the RPS, we point out an analogy to the Ising model. It can be verified that the two-particle version of our process (i.e.~ $AB \uto{1} AA$, $BA \uto{1} BB$ and $A \uto{\mu} B \uto{\mu} A $; $\mu_l$ and $\mu_r$ obviously cannot be distinguished here) has been proposed by Glauber as a way to study the dynamic effects of the one-dimensional Ising model, with, say, $A$ corresponding to ``spin up'' and $B$ corresponding to ``spin down'' \cite{glauber,schuetz}. After expressing the energy in terms of the nearest neighbor sum $E(\{s\}) = - J \sum s_k s_l$, where $s_k$ is 1 for ``spin up'' and $-1$ for ``spin down'' and $J$ is a coupling constant, the temperature $T$ is related to the mutation rate $\mu$ by
$ \mu/(1+\mu) = 1 - \tanh(2kJ/T)$.
$\mu$ is small in the low temperature regime and large in the high temperature regime. Thus, we may think of the mutation rates $\mu_l$ and $\mu_r$ for the RPS as temperature-like parameters. At fast mutation rates, or high temperature, the system becomes rather uncorrelated, and therefore mean-field (\ref{eq:mean-field}) makes for a good approximation. 

Let us now try to comprehend the RPS for very low mutation rates $\mu = \mu_l + \mu_r$. Comparison with the one-dimensional Ising model suggests that at $\mu=0$, corresponding to zero temperature, the system displays a critical behavior with the correlation length going to infinity. In the following we show that this is indeed corroborated by scaling arguments as well as stochastic simulations. 

\begin{figure}
      \centering \includegraphics[width=0.37\textwidth]{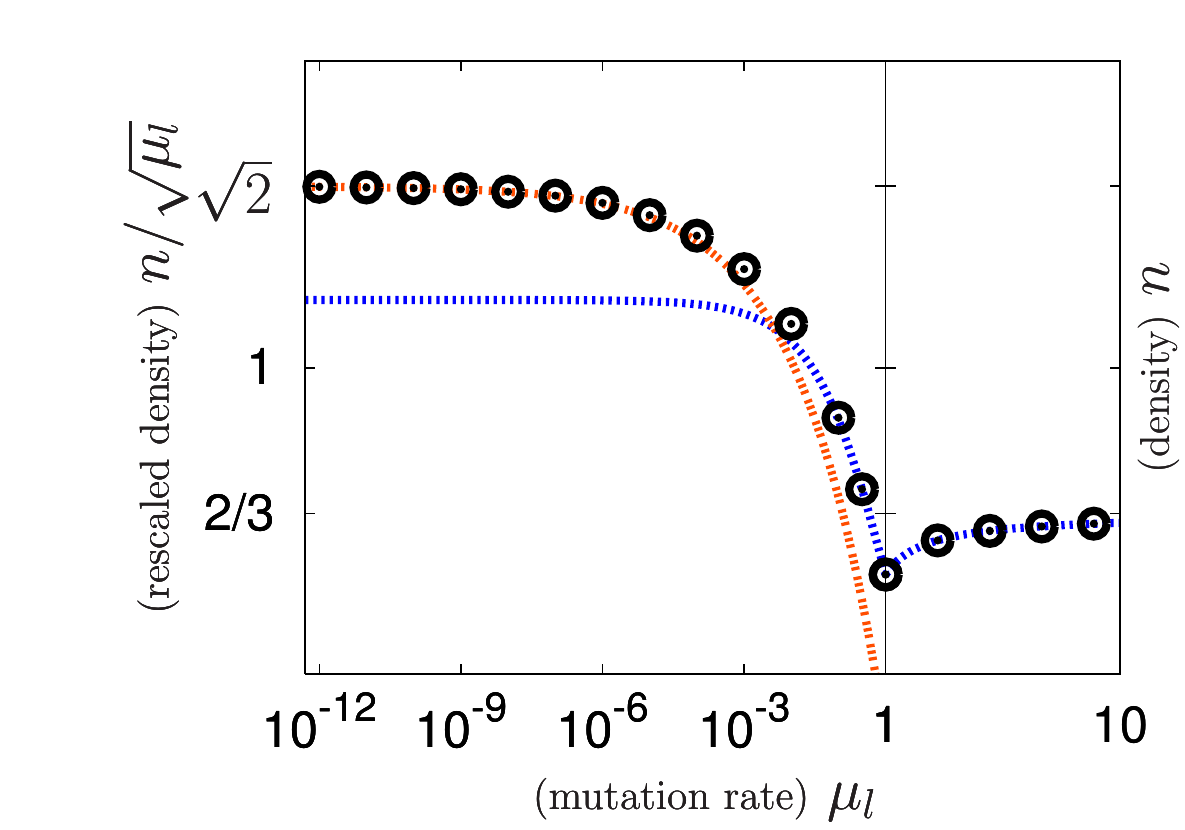}
\caption{When $r_A,r_B ,r_C=1$,  the rescaled interface density $n(\mu_l, \mu_r = 0)/\sqrt{\mu_l}$ approaches the law $\sqrt{2} -  (3/4) 2^{3/4} \mu_l^{1/4}$ (red line) as $\mu_l$ becomes small. For large mutation rates the data is described well by the mean-field curve (\ref{eq:mean-field}) (blue line). Notice that for $\mu_l > 1$ we have plotted the density $n$ without rescaling.
}
\label{fig:rpsscaleml}
\end{figure}

First, let us restrict ourselves to the regime $\mu_r = 0$ and $\mu_l \ll 1$. The interface density is low and therefore the single most probable mutation occurs on two adjacent, vacant sites
  $\oo \oo \uto{\mu_l} LR$.
In the particle picture this is achieved by, e.g.~, $AAA \uto{\mu_l} ACA$. Notice that the mutation induces a predator in a---typically large---domain of prey, where it can spread subsequently. Hence the incidence has strong impact on the system. In the dual picture this is expressed in the fact that the pair $LR$, unlike $RL$, can separate, e.g.
\begin{equation}
 \oo \oo \oo \oo \uto{\mu_l} \oo LR \oo  \uto{1} L \oo R \oo \uto{1} L \oo \oo R \uto{\phantom{1}} \ldots \,.
\end{equation}
Consider what happens next to the, say, $R$ interface. It moves to the right from site to site with rate $1$, until it meets and reacts with some other interface, e.g., 
\begin{equation}
  R \oo \oo L \uto{1} \oo R \oo L \uto{1} \oo RL \oo \uto{2} \oo \oo\oo \oo \,.
\end{equation}
It is crucial to note that diffusion becomes negligible when the particles are far apart, since their directional motion is described by a Poisson process, whose mean square displacement $\sigma (t) = \sqrt{t}$ becomes small relative to the average distance $\langle x(t) \rangle = t$ it has traveled. Therefore, in our regime one should think of the particles as moving ballistically. For our scaling argument,  we partition the lattice in cells of size $b$ and consider the dynamics from this coarse-grained point of view. Empty cells become the new vacancies $\oo$, cells that contain exactly one interface the new $R$ or $L$, respectively. Since the lattice is supposed to be sparsely populated, we disregard the unlikely case of cells containing more than one interface.
A mutation $\mu_l$ now occurs with $b$-fold rate, since the whole cell is at disposal. We rescale time by a factor of $b$, so that the velocity of the (ballistic) interfaces is unchanged. This implies a rescaled rate $\mu_{l}' = b^2 \mu_{l}$---one factor $b$ for rescaling space and another one for rescaling time. 
The density evidently becomes $b$-fold, $n( \mu_{l}') = b n(\mu_l)$ and thus, $n = \mathscr{A} \mu_{l}^{1/2}$ for an infinite lattice. This result is indeed validated by our numerical simulations (Figure~\ref{fig:rpsscaleml}).

At this point, we remark that for the symmetric case $r_A, r_B, r_C =1 $ the stationary state can be solved exactly, leading to $\mathscr{A} = \sqrt{2}$ \cite{winkler-2010}.
Here we only illustrate the underlying physics by the following heuristic argument. For symmetric rates, reactions of the type $RR \to \oo L$ are negligible, because it takes an interface much longer to catch up with an interface of the same kind than to crash into a different kind of interface, which it can meet halfway, as it were. Hence $P(RR) = 0$, where $P(RR)$ stands for the probability of finding two interfaces next to each other. We suppose that otherwise the system is uncorrelated, in particular $P(RL) = P(R) P(L)$. Then, up to terms of the order $\mu_l$ and due to the symmetry between $R$ and $L$ one has the master equation, 
\begin{equation}
 \dot{P}(R) \approx \mu_l P(\oo\oo) - 2 P(RL) \approx \mu_l - 2 \left[P(R)\right]^2 \,.
\label{eq:heur}
\end{equation}
Solving for the stationary value yields $\mathscr{A} = \sqrt{2}$.

To motivate the first correction to this result, notice that two interfaces of the same kind move diffusively relative to each other, with diffusion constant $1$. The diffusional lengthscale associated to the average survival time $1/n$ is $1/\sqrt{n}\ll 1/n$ (for small mutation rates). The probability $P_n(t)$ that the pair of $R$-interfaces is intact after a time $t$ will just be the probability that they have not yet interacted diffusively, $\frac{2}{\sqrt{\pi}} \int_0^{x/2\sqrt{t}} \! ds \;  e^{-s^2}$ (see, e.g., \cite{redner}), times the probability that the right $R$ has not yet crashed into an $L$, which is given by an exponential distribution $\exp\left(- n t\right)$. In the last equation we assume that the system is uncorrelated, so that in every time step the right $R$ crashes into an $L$ with probability $n$. The probability that an $R$ is created at a distance $x$ to another $R$ is n/2 when $x \ll 1/n$. Upon integrating over $x$ and $t$ one finds that the probability of an $R$ interacting diffusively with an $R$ on its right is $\sqrt{n}/2$. If we simply subtract these "failed attempts" of creating it from the mutation rate $\mu_l \to \mu_l(1 - \sqrt{n}/2)$, we obtain $\mu_l (1-n/2) \cdot 1/n = n/2$ or $n \approx \sqrt{2 \mu_l} (1 - \sqrt{n}/4) \approx  \sqrt{2 \mu_l} - \left(2^{3/4}/4\right) \mu_l^{3/4}$. More rigorous analysis, taking into account the difference in the time of survival of the left $R$s, with and without diffusion, yields an even larger correction 
\begin{equation}
	n \approx \sqrt{2 \mu_l} - \frac{3}{4} 2^{3/4} \mu_l^{3/4}\,,
	\label{eq:mul}
\end{equation}
(see supplementary material). 
Our result is in agreement with the numerics (Figure~\ref{fig:rpsscaleml}). 

\begin{figure}
      \centering \includegraphics[width=0.37\textwidth]{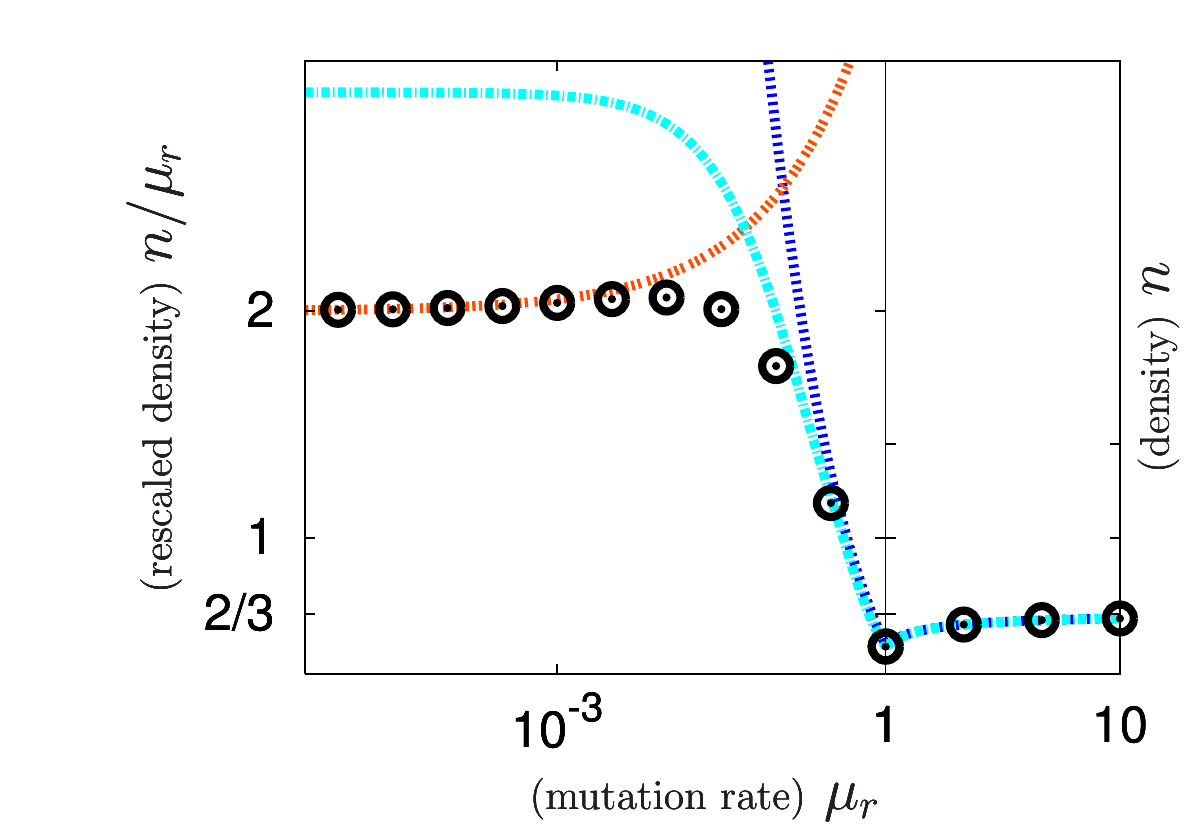}
\caption{For small $\mu_r$, the rescaled density converges to $n(\mu_l = 0,\mu_r)/\mu_r= 2+3/2 \sqrt{\mu_r}$ (red line), when $r_A, r_B, r_C = 1$. The mean-field result (blue line) and the generalized mean-field result for clusters containing two sites (turquoise line) is also shown. 
}
\label{fig:mur}
\end{figure}

Let us proceed to discuss the case $\mu_l = 0$ and $\mu_r \ll 1$. Since the system is coarse grained, the major part of the mutations will result in one prey in the middle of large domains of predators. For instance, $AAA \uto{\mu_r} ABA$. Evidently, this configuration is rather unstable and one expects that in most cases the cyclic dominance reactions reestablish the original state, that is $B$ turns to $A$. In the dual picture, this translates into the creation of a pair $RL$, which in most cases annihilates quickly,
\begin{equation}
 \oo\oo {\ }^{\underrightarrow{\ \mu_r \ }} RL {\ }^{\underrightarrow{\ 2 \ }} \oo\oo \,.
 \label{eq:mur1}
\end{equation}
Owing to the longevity of its products, one also needs to take into account that a second mutation may occur,
\begin{equation}
 \oo\oo {\ }^{\underrightarrow{\ \mu_r \ }} RL {\ }^{\underrightarrow{\ \mu_r \ }} LR \,,
 \label{eq:mur2}
\end{equation}
effectively leading to $\oo\oo {\ }^{\underrightarrow{\ \mu_r^2/2 \ }} LR$.
Just as above a pair $LR$ is produced, but this time the reaction is mediated by two $\mu_r$ mutations instead of one $\mu_l$ mutation. In the particle picture this means that the prey $B$ in a domain of $A$ may be turned into the predator $C$ by a second mutation. The former (\ref{eq:mur1}) implies a contribution of $\mu_r$ to the interface density. The latter (\ref{eq:mur2}) leads to the same dynamics as in the $\mu_l$ case and to another term of magnitude $\sqrt{2 \left(\mu_r^2/2\right)} = \mu_r$, i.e. to lowest order $n = 2 \mu_r$.
 
For the leading correction, in addition to reactions of the type $RR\to \oo L$, one needs to treat instances of mutations when there exactly one interface around, say
\begin{equation}
 R \oo \uto{\mu_r} LL\,,
 \label{eq:mur3}
\end{equation}
e.g., $ABB \uto{\mu_r} ACB$.
Similar reactions can occur, when there is a mutation nearby an interface, for instance, 
\be
\label{eq:mur8}
 R \oo \oo \oo  \uto{\mu_r}  R \oo R L \uto{1} \oo R R L \uto{1} \oo \oo L L \,.
\ee
An analysis analogous to the pure $\mu_l$ case yields an overall positive contribution \cite{winkler-2010},
\begin{equation}
 n \approx 2 \mu_r + \frac{3}{2} \mu_r^{3/2}\,.
 \label{eq:mur}
\end{equation}
Figure~\ref{fig:mur} confirms this behavior. Again mean-field is an excellent approximation for large mutation rates. As mutations become less frequent, equation (\ref{eq:mean-field}) provides a gross over-estimate of the interface density, because the approach cannot keep track of the large amount of pairs of $RL$ that annihilate quickly. This can be amended by a generalized mean-field approach \cite{szabo-2007-446}, where the master equation for  clusters of $N$ adjacent sites is considered. A truncation in the hierarchy of probability distributions yields a closed set of differential equations, which can be solved numerically. For clusters of size 2 one already retrieves the right scaling law $n \sim \mu_r$ ($\mu_r \ll 1$).

Again, we explain how a scaling analysis helps us understand the behavior of the density $n(\mu_r,\mu_l=0)$. This will also extend our results to more general processes. We partition the lattice in cells of size $b$. Then the probability for a cell to contain a pair $RL$, which are created and destroyed according to reaction (\ref{eq:mur1}), becomes $b \mu_r$ while the rate to create a pair $LR$ out of $RL$ (\ref{eq:mur2}) evidently remains $\mu_r$. We rescale time by a factor $b$ so that the velocity of $R$ and $L$, measured in the average number of cells they traverse in unit time, stays one. Now the right-hand side of reaction~(\ref{eq:mur2}) occurs at a rate $b \mu_r$, while the probability of finding a pair $LR$ remains $b \mu_r$. This implies $\mu_r'=b \mu_r, n(\mu_r')= b n$, whereby we conclude $n = \mathscr{B} \mu_r$. 

\begin{figure}
      \centering \includegraphics[width=0.37\textwidth]{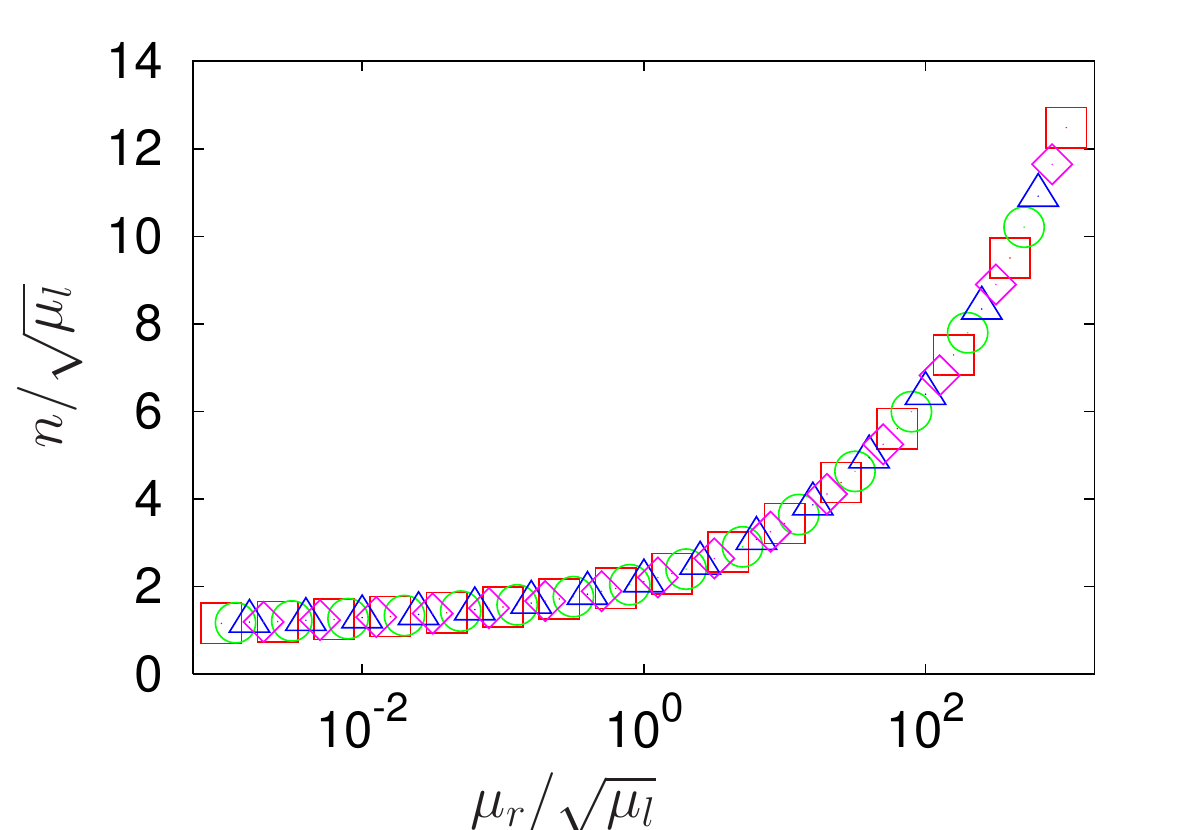}
    \caption{Collapse of the data onto a scaling function.
     It is demonstrated that our arguments hold true also for asymmetric rates, here $r_A = 5$, $r_B = r_C = 1$. The data points are $\mu_l = 2^{-10} (\mbox{\textbf{\footnotesize{\textcolor{red}{$\square$}}}}) ,2^{-11} (\mbox{\Large{\textcolor{green}{$\circ$}}}) ,2^{-12} (\mbox{\textbf{\footnotesize{\textcolor{blue}{$\bigtriangleup$}}}}), 2^{-13} (\mbox{\Large{\textcolor{magenta}{$\diamond$}}})$. 
 }
\label{fig:rpsscale}
\end{figure}

Up to now we have considered symmetric rates, where reactions governed by diffusion  
give rise to the correction terms in (\ref{eq:mul},\ref{eq:mur})
which do not fit in the scaling.
Now consider what happens if $r_A,r_B,r_C$ are not identical, for instance if $r_A > r_B = r_C$. Suppose the pair $RR$ stands for $CAB$. In this case the two $Rs$ no longer move diffusively relative to each other but rather the right $R$ drifts away and can escape the left one,  resembling a ballistic motion of the right $R$ away from the left one with average relative velocity $r_A - r_B$. Here we can apply the above scaling argument and we expect a contribution to the interface density that obeys the scaling $n \sim \mu_r$ as confirmed in Figure~\ref{fig:rpsscale}. This plot also indicates that our findings can be generalized to $\mu_l \ne 0$ and $\mu_r \ne 0$, where in analogy to our above arguments one derives $n(\mu_l,\mu_r) = \sqrt{\mu_l} \, \phi \! \left(\frac{\mu_r}{\sqrt{\mu_l}} \right)$ for some scaling function $\phi$. We remark that exact calculations yield $\phi \! \left( 0 \right) {\ }^{\underrightarrow{\ r_A \to \infty \ }} 1$. Figure~\ref{fig:rpsscale} shows that even for $r_A = 5$ this is a good approximation.

In conclusion, competition between coarsening dynamics and mutations in our model leads to a reactive stationary state characterized by an interplay between equilibrium and non-equilibrium processes---indeed one can pinpoint exactly which reactions take the system away from equilibrium. It was crucial to discriminate two types of mutations, $\mu_l$ and $\mu_r$,  the effect of the latter being negligible when the two rates are comparable and small. Both for the high and for the low mutation rate regime we have retrieved asymptotically exact results which we expect to be quite robust to a wide rage of variations, e.g.~relaxing the constraints of perfect symmetry.

Financial support of the German Excellence Initiative via the program ``Nanosystems Initiative Munich'' and the German Research Foundation via the SFB TR12 ``Symmetries and Universalities in Mesoscopic Systems'' is gratefully acknowledged.

\onecolumngrid
\newpage

\begin{center} {\LARGE \textbf{ Supplementary Material} (EPAPS Document)}   \end{center}

\section*{CALCULATION OF THE LEADING CORRECTION TO THE DENSITY}

In this supplement we show how to calculate the leading correction to the law $n = \sqrt{2 \mu_l}$ for the interface density, when $\mu_l \ll 1$ and $\mu_r = 0$. 
Let $\tn$ denote the density of interfaces that will ultimately annihilate ballistically, i.e. by pair annihilation $RL \to  \oo \oo$. We expect that $\tn \approx \sqrt{2 \mu_l}$. In a first step, we calculate the ratio of, say, $R$s that annihilate diffusively, i.e. via the reaction $RR \to \oo L$. 
Suppose an $R$ is                                                                                                                                                                                                                                                                                                                                                                                                                                                                                                         created $x$ sites to the left of the nearest further $R$ to the right (there may only be empty sites and $L$s in between them). For instance, if there are no $L$s in between, which is the most important case, this looks like
\[\underbrace{R\oo\ldots\oo}_{x \mbox{ sites}}R\,. \]
Relative to each other the two interfaces move diffusively, with diffusion constant $1$, and thus the two $R$s might crash into each other through this diffusional motion. 
The expected lifetime is of the order $1/n$, so that the relevant  diffusional distances  are of the order $1/\sqrt{n}$. Therefore, we may assume $x\ll 1/n$ in the following and neglect the possibility of an $L$ in between the two interfaces. 

The probability $P_n(t)$ that the pair of $R$-interfaces is intact after a time $t$ equals the probability that they have not yet interacted diffusively, $P_{d}(x,t) = \frac{2}{\sqrt{\pi}} \int_0^{x/2\sqrt{t}} \! ds \;  e^{-s^2}$, times the probability that the right $R$ has not yet crashed into an $L$, which is equal to $P_{b}(t) = \exp\left(- \tn t\right)$. In the last equation we assume that the system is uncorrelated, so that in every time step the right $R$ crashes into an $L$ with probability $\tn$, giving rise to an exponential distribution with parameter $\tn$. 
Furthermore, we note that when $x \ll 1/n$ the probability that an interface is created at $x$ is $n/2 \approx \tn/2$ (since the density of $R$ is $n/2$). Thus the probability that a particle annihilates diffusively becomes
\[
	 - \int_{0}^{c} dx \int_{0}^{\infty} dt \,  \frac{\tn}{2} \left(\frac{\partial}{\partial t} P_d(x,t)\right) P_b(t) \approx - \int_{0}^{\infty} dx \int_{0}^{\infty} dt \,  \frac{\tn}{2} \left(\frac{\partial}{\partial t} P_d(x,t)\right) P_b(t) = \frac{\sqrt{\tn}}{2}\,,
\]
where $1/\sqrt{n} \ll c \ll 1/n$.
For uncorrelated interfaces, the average time until a particle annihilates ballistically is $1/\tn$. Multiplying the input rate of the interfaces, $2 \mu_l$, by the probability that there is no diffusive interaction, $1-\sqrt{\tn}/2$, 
the density of particles that annihilate ballistically becomes $\tn =   \left[2 \mu_l \left(1-\sqrt{\tn}/2\right) \right] \frac{1}{\tn}$, or
\[
	\tn =  \sqrt{2 \mu_l \left(1 - \frac{\sqrt{\tn}}{2}\right)} \approx \sqrt{2 \mu_l} \left( 1 - \frac{\sqrt{\tn}}{4}\right) \approx \sqrt{2 \mu_l} - \frac{2^{3/4}}{4} \mu_l^{3/4}\,.
\]
$\tn$ is not the interface density that we are looking for. To find $n$ we need to add the contribution of the particles that annihilate diffusively. To do this we look at the average time of survival $\tau(x)$ of an interface that is created at $x$. We also introduce the quantity $\tau_b(x)$, which denotes the average time of survival for truly ballistic interfaces. 
We know that if diffusion can be neglected, the average time of survival is $1/\tn$, i.e.
\[
	\langle \tau_b \rangle \equiv \int_0^\infty dx\, \rho(x) \tau_b(x) = \frac{1}{\tn}\,,
\]
where $\rho(x)$ denotes the probability that an $R$ is created at $x$.

Without diffusion, the average time of survival of an $R$ that is created at $x \ll 1/n$ would be $2 \cdot 1/\tn$ since the right $R$ must be annihilated before the left one can be destroyed. Together with diffusion one has instead
\[
	\tau(x) = - \left[ \int_0^\infty dt\,  t \left(\frac{\partial}{\partial t} P_{d}\right) P_{b} + (t + 1/\tn) P_{d} \left( \frac{\partial}{\partial t} P_{b} \right) \right] \,.
\]
(Notice that upon ballistic annihilation of the right $R$ at time $t$, the left one lives on for $1/\tn$ units of time on average, whence the factor $t + 1/\tn$.)
For $x \gg 1/\sqrt{n}$ diffusion is negligible, $\tau(x) \equiv \tau_b(x)$, justifying the boundary $c$ in the following integral. Again we remark that when $x \ll 1/n$ then $\rho(x) = n/2 \approx \tn/2$. Thus
\[
	\langle \tau \rangle - \langle \tau_b \rangle \approx \int_0^c dx\, \frac{\tn}{2} \left(\tau(x) - \frac{2}{\tn}\right) \approx \int_0^\infty dx\, \frac{\tn}{2} \left(\tau(x) - \frac{2}{\tn}\right) = - \frac{1}{\sqrt{\tn}} \,,
\]
where $1/\sqrt{n} \ll c \ll 1/n$. Therefore,
\[
	\langle \tau \rangle = \frac{1}{\tn} - \frac{1}{\sqrt{\tn}}\,.
\]
Expressing $\tn$ in terms of $\mu_l$ we find for the interface density
\[
	n = 2 \mu_l \langle \tau \rangle \approx \sqrt{2 \mu_l} - \frac{3}{4} 2^{3/4} \mu_l^{3/4}\,.
\]


\begin{thebibliography}{17}
\bibitem[{win()}]{winkler-2010}
\bibinfo{note}{A detailed analysis is given in a forthcoming publication.}
\end{thebibliography}
\end{document}